\documentclass[10pt,a4paper,final]{iopart}
\usepackage{iopams}
\usepackage{graphicx}
\usepackage[breaklinks=true,colorlinks=true,linkcolor=blue,urlcolor=blue,citecolor=blue]{hyperref}
\usepackage{subcaption}
\usepackage{bbm}
\usepackage{xcolor}
\expandafter\let\csname equation*\endcsname\relax
\expandafter\let\csname endequation*\endcsname\relax
\usepackage{amsmath,amsfonts,amssymb,amsthm}
\usepackage{braket}

\newcommand{\id}{\mathbbm{1}}

\newtheorem{prop}{Proposition}

\newcommand{\be}{\begin{equation}}
\newcommand{\ee}{\end{equation}}

\begin{document}
\title[Quantum statistical models with parameter dependent rank]{On the discontinuity of the quantum Fisher information for 
quantum statistical models with parameter dependent rank}
\author{Luigi Seveso}
\address{Quantum Technology Lab, Dipartimento di Fisica ``Aldo Pontremoli'', Universit\`a degli Studi di Milano, I-20133 Milano, Italy}
\author{Francesco Albarelli}
\address{Department of Physics, University of Warwick, Coventry 
CV4 7AL, United Kingdom}
\author{Marco G. Genoni}
\address{Quantum Technology Lab, Dipartimento di Fisica ``Aldo Pontremoli'', 
Universit\`a degli Studi di Milano, I-20133 Milano, Italy}
\author{Matteo G. A. Paris$^{1,2}$}
\address{$^1$Quantum Technology Lab, Dipartimento di Fisica ``Aldo Pontremoli'', 
Universit\`a degli Studi di Milano, I-20133 Milano, Italy}
\address{$^2$INFN, Sezione di Milano, I-20133 Milano, Italy}
\date{\today}
\begin{abstract}
We address the discontinuities of the quantum Fisher information 
(QFI) that may arise when the parameter of interest takes values
that change the rank of the quantum statistical model. We revisit 
the classical and the quantum Cram\'er-Rao theorems, show that 
they do not hold in these limiting cases, and discuss how this 
impacts on the relationship between the QFI and the Bures metric.
In order to illustrate the metrological implications of
our findings, we present two paradigmatic examples, where we
discuss in detail the role of the discontinuities. 
We show that the usual equivalence between the variance of the maximum likelihood estimator and inverse QFI breaks down.
\end{abstract}
\maketitle
\section{Introduction}
A quantum metrological protocol is a detection
scheme where the inherent fragility of quantum systems to external 
perturbations is exploited to enhance precision, stability or resolution 
in the estimation of one or more quantities of interest. In 
the last two decades, the development of advanced technologies 
to coherently manipulate quantum systems, and to address them 
with unprecedented accuracy, made it possible 
to realize several metrological schemes based on quantum systems,
leading to quantum enhanced high-precision 
measurements of physical parameters~\cite{Giovannetti2011}.

On the theoretical side, the main tool of quantum metrology
is the so-called quantum Cram\'er-Rao theorem, stating that for
a regular quantum statistical model the precision is bounded
by the inverse of the quantum Fisher information (QFI)~\cite{helstrom1976quantum,Holevo2011b,Hayashi2005,Braunstein1994,Paris2009,Petz2010}. 
Evaluating the QFI thus provides the ultimate quantum limits
to precision, and a general benchmark to assess metrological 
protocols.
The quantum Cram\'er-Rao theorem is indeed a very powerful 
tool, and it has found a widespread use in quantum metrology.
At the same time, its success has lead to somehow overlooking 
the mathematical details of its hypotheses, such as a possible intrinsic parameter dependence of the measurement apparatus~\cite{Seveso2016a,Seveso2018a} or the pathological situations that may occur when the parameter of interest takes values that change the rank of density matrix of the system.
In such quantum statistical models, analogous to \emph{non-regular models} in classical statistics, the QFI may 
show discontinuities, which undermine the validity of the 
Cram\'er Rao theorem and, in turn, its use in quantum metrology.

In this paper, we consider statistical models whose 
rank is a non-trivial function of the parameter to be 
estimated. We address the discontinuities of classical and 
quantum Fisher information and revisit both the classical 
and the quantum Cram\'er-Rao theorems, showing that 
they do not hold in these limiting cases, also discussing 
how this reflects on the relationship between the QFI 
and the Bures metric.
In order to illustrate the metrological implications of
our findings, we also discuss two paradigmatic simple examples, where 
the Cram\'er-Rao bound (CRB) may be easily violated.

Let $\rho_{\theta}$ 
denote a quantum statistical model with parameter space $\Theta$. Suppose 
that $\bar \theta$ is the true value of the parameter and that, in any open neighbourhood $N_{\bar \theta}$ of $\bar \theta$, there exists $\theta'$ 
such that $\text{rank}(\rho_{\theta'})\neq \text{rank}(\rho_{\bar \theta})$. 
The typical situation is when the rank changes at an isolated point 
$\bar \theta$ of $\Theta$, but more general situations may also be envisioned.
This apparently harmless circumstance causes new theoretical 
challenges in determining the best performance of any quantum estimation 
strategy.
Nonetheless, it is a situation of physical interest that might naturally arise when estimating noise parameters, e.g. in the estimation of momentum diffusion induced by collapse models under continuous monitoring of the environment~\cite{Genoni2016a}.
We will show that this scenario also applies to an instance of frequency estimation with open quantum systems~\cite{Haase2018}.

The consequences of allowing the rank to vary with $\theta$ can be severe, 
both from a geometrical and a statistical perspective.
From the geometrical point of view, it is known that the Fisher information metric may develop discontinuities~\cite{Safranek2017a} and suitable regularization techniques have been proposed for specific classes of states~\cite{Safranek2019,serafini2017quantum}.
On the other hand, the question of how such discontinuities affect the statistical estimation problem at hand is currently open~\cite{Safranek2017a}.
In the following, we are going to argue that the standard theory based 
on the Cram\'er-Rao bound  breaks down at such points of the parameter space 
where the rank of $\rho_\theta$ changes.
In fact, such a failure of the standard theory is not specific to \emph{quantum} statistical models, but is actually present already at the \emph{classical} level~\cite{cramer1946mathematical}; a generalized classical CRB for this scenario has been recently derived~\cite{Bar-Shalom2014,Lu2017b}.

\subsection{Classical and quantum regular models}
In order to establish notation, let us briefly review the regular scenario, 
which is the theoretical foundation to most applications in classical and 
quantum metrology. We assume that the quantum parametrization maps 
$\varphi_\theta:\,\theta\to \rho_\theta$ or the classical one
$\varphi_\theta:\,\theta\to p_\theta$ are sufficiently well-behaved, such 
that  the symmetric logarithmic derivative $L_\theta$, implicitly defined via the relation $\partial_\theta \rho_\theta = \{\rho_\theta,\,L_\theta\}/2$ or 
the classical score function $\ell_\theta = \partial_\theta \log p_\theta$ 
exist and the corresponding quantum and classical Fisher information metrics are  well-defined and finite, $\forall \theta \in \Theta$. 

Setting apart all 
pathological situations where these conditions do not hold, e.g. the statistical model is non-differentiable or even discontinuous (see~\cite{Tsuda2005,Yang2018a} for such a scenario in quantum estimation and~\cite{Akahira1995} for classical estimation), we further qualify a quantum statistical model as \emph{regular} if it satisfies the following conditions: 
1. \emph{fixed-rank}: the rank of the statistical model 
$\rho_\theta$ (i.e. the rank of the density matrices $\rho_\theta$) 
is independent of $\theta$; 2. \emph{identifiable}: the parametrization 
map $\varphi_\theta:\,\theta\to \rho_\theta$ is injective; 3.
\emph{non-singular metric}: the Fisher-Bures metric $g_\theta$, defined
by $2  [1-F(\rho_\theta,\rho_{\theta+\epsilon})] = g_\theta\, \epsilon^2 + O(\epsilon^3)$, where $F(\rho,\sigma)=
\tr \left[\sqrt{\sqrt{\rho}\sigma\sqrt{\rho}}\right]$ is the quantum fidelity,
is a well-defined positive-definite function $\forall  \theta \in \Theta$.
\par
Let us also give the translation of the previous definition to the classical 
setting. A \emph{regular classical statistical model} $p_\theta$ satisfies 
the following conditions: 1. \emph{parameter-independent support}: the 
support $\text{supp}(p_\theta)$ of the statistical model (i.e. the subset 
of the real axis where $p_\theta\neq 0$) is independent of $\theta$\footnote{This is one of the conditions needed to make sure that the order of differentiation with respect to $\theta$ and integration over the sample space can be interchanged, for more mathematical details see~\cite[p.~516]{casella2002statistical} and also~\cite{Bar-Shalom2014} and references therein};
2. \emph{identifiable}: the coordinate map $\varphi_\theta:\,\theta\to 
p_\theta$ is injective; 3. \emph{non-singular metric}: the Fisher-Rao 
information metric $f_\theta$, defined by $2 D(p_\theta||p_{\theta+\epsilon}) 
= f_\theta\, \epsilon^2 + O(\epsilon^3)$, where $D(p||q) = \sum_x p_x \log p_x/q_x$ 
is the Kullback-Leibler divergence, is a well-defined positive-definite 
function $\forall  \theta \in \Theta$.

For regular statistical models one has the following results~, which provide the standard tools of current classical and quantum metrology
\begin{prop}\label{crprop1}
Given a regular classical statistical model $p_\theta$,
\begin{itemize}
\item  For any unbiased estimator $\hat \theta$, the Cram\'er-Rao bound $\text{Var}_\theta(\hat \theta)\geq \left(M\,\mathcal F_\theta\right)^{-1}$ holds, where $M$ is the number of repetitions and $\mathcal F_\theta$ the Fisher information, 
\be
\mathcal{F}_\theta = \sum_y \frac{\left[\partial_\theta p_\theta(y)\right]^2}{p_\theta(y)}
\ee
\item The Cram\'er-Rao bound is attainable: \emph{i)} for finite $M$, if $p_\theta$ belongs to the exponential family and $\theta$ is a natural parameter of $p_\theta$, by the unique efficient estimator of $\theta$ \emph{ii)} asymptotically, as $M\to\infty$, e.g. by the maximum-likelihood or Bayes estimators.   
\item The maximum-likelihood and Bayes estimators have the asymptotic normality property, i.e. convergence in distribution $\sqrt M (\hat \theta - \theta)\overset{d}{\to}\,N(0,1/\mathcal F_\theta)$ as $M\to \infty$. 
\end{itemize}
\end{prop}
In addition, we have that $\mathcal{F}_\theta=f_\theta$ i.e the Fisher information equals the Fisher-Rao metric.
\begin{prop}\label{crprop2}
Given a regular quantum statistical model $\rho_\theta$,
\begin{itemize}
\item  For any quantum measurement $\{\Pi_x\}_{x\in\mathcal X}$, $\sum_{x\in\mathcal X} \Pi_x = \mathbbm I$ and unbiased estimator $\hat \theta$, the quantum Cram\'er-Rao bound $\text{Var}_\theta(\hat \theta)\geq \left(M\, \mathcal{Q}_\theta\right)^{-1}$ holds, where $M$ is the number of repetitions, $\mathcal{Q}_\theta$ the quantum Fisher information,
\begin{align}
\label{eq:QFIdef1}
\mathcal{Q}_\theta = \tr[ \rho_\theta L_\theta^2] \,
\end{align}
and $L_\theta$ is the symmetric logarithmic derivative (SLD) operator, defined via the Lyapunov equation $2 \partial_\theta \rho_\theta = L_\theta \rho_\theta + \rho_\theta L_\theta$.
\item The quantum Cram\'er-Rao bound is attainable by implementing the optimal Braunstein-Caves measurement, i.e. a projective measurement of the symmetric logarithmic derivative $L_\theta$, and under the conditions stated before for the resulting classical statistical model.
\end{itemize}
\end{prop}
In the quantum setting the optimal measurement generally depends on the true value of the parameter $\theta$.
Nonetheless, the single-parameter quantum CRB is attainable in the limit of many repetitions by implementing an adaptive strategy~\cite{Nagaoka1989a,Fujiwara2006}, such as a two-stage adaptive measurement~\cite{Hayashi1998a,Barndorff-Nielsen2000}.

By considering the spectral decomposition of the quantum statistical model $\rho_\theta = \sum_k \lambda_{k,\theta} | \lambda_{k,\theta} \rangle\langle \lambda_{k,\theta} |$, one obtains that the SLD operator can be written as~\cite{Paris2009}
\begin{align}
L_\theta = 2 \sum_{\tiny{\lambda_{k,\theta} + \lambda_{l,\theta} > 0}} \frac{\langle \lambda_{k,\theta} | \partial_\theta \rho_\theta | \lambda_{l,\theta} \rangle}{\lambda_{k,\theta} + \lambda_{l,\theta}} |\lambda_{k,\theta} \rangle \langle \lambda_{l,\theta} | \,, 
\end{align}
and consequently the QFI can be evaluated as
\begin{equation}
\mathcal{Q}_\theta = 2 \sum_{\tiny{\lambda_{k,\theta} + \lambda_{l,\theta} > 0}} \frac{ |\langle \lambda_{k,\theta} | \partial_\theta \rho_\theta | \lambda_{l,\theta} \rangle | ^2 } {\lambda_{k,\theta} + \lambda_{l,\theta}} \,. \label{eq:formulaQFI}
\end{equation}
Moreover, the QFI is proportional to the Fisher-Bures metric
\begin{align}
\mathcal{Q}_\theta = 4 g_\theta\,.
\end{align}
We remark that for non full-rank quantum models the Lyapunov equation does not have a unique solution.
Nonetheless, there is no ambiguity in the definition of the QFI for fixed-rank models, since the unspecified components of the SLD do not play any role in Eq.~\eqref{eq:QFIdef1}~\cite{Holevo2011b,Liu2014a}.
\section{Non-regular case}
If either of the three regularity conditions listed above is not true, Props.~\ref{crprop1} and~\ref{crprop2} do not hold in general.
The second condition, if not verified, can often be realized by simply restricting the parameter space $\Theta$ or by a change of parametrization.
Let us consider a trivial example, i.e. the statistical model $\rho_\theta = \sin^2\theta \ket{0}\bra{0} + \cos^2\theta \ket{1}\bra{1}$.
If we consider $\theta \in \mathbbm R$, the model is not identifiable, but it can be made so by restricting the values of $\theta$ to the interval $[0,\pi/2]$.
In a non-identifiable model, the true value of the parameter is in general non-unique, therefore a local approach becomes impossible and the Cram\'er-Rao bound is meaningless.  

\par
Let us now assume that the model is identifiable, or can be made so by 
a suitable reparametrization. However, the rank of the statistical 
model, or its support in the classical case, is allowed to vary by 
varying the parameter $\theta$. 
\par
\subsection{Variable-rank models}
Let us start from the classical case, since our conclusions may then be 
translated to the quantum case.
Denote by $\mathcal X_\theta$ the support of $p_\theta$, i.e. the closure of the set $\{x\,|\, p_\theta(x)>0\}$.
Let's see why the derivation of the Cram\'er-Rao bound breaks down. For simplicity, fix $M=1$.
Given any two statistics $t_1$ and $t_2$, their inner product is defined in terms of the probability distribution $p_\theta(x)$ as
\be
\langle t_1,\,t_2\rangle = \mathbbm E_\theta(t_1 t_2)=\int_{\mathcal X_\theta} dx\, p_\theta(x) t_1(x) t_2(x)\;.
\ee
Take $t_1(x)= \hat \theta(x)-\theta$ and $t_2(x)=\partial_\theta \log p_\theta(x)$. Then, by the Cauchy-Schwarz inequality $\langle t_1,\,t_2\rangle^2 \leq \langle t_1,\,t_1\rangle\langle t_2,\,t_2\rangle$, one obtains
\be
\int_{\mathcal X_\theta} dx\, [\hat \theta(x)-\theta]\, \partial_\theta p_\theta(x) \leq \text{Var}_\theta(\hat \theta)\cdot \mathcal F_\theta\;.
\ee
Now, if $\mathcal X_\theta$ were independent of $\theta$, and under very mild assumptions regarding the smoothness of $p_\theta$ (see e.g.~\cite[p.~516]{casella2002statistical}), one could interchange the order of integration and differentiation, and conclude that the LHS is equal to 1, which would imply the Cram\'er-Rao bound.
However, the very fact that $\mathcal X_\theta$ depends on $\theta$ prevents one from interchanging integration and differentiation and thus to obtain a general inequality independent from the particular unbiased estimator $\hat \theta$. 
The conclusion is that in this situation the Fisher information is not necessarily linked to the best possible precision of unbiased estimators.

Moving to the quantum case, since the quantum Fisher information is the Fisher information corresponding to the optimal measurement~\cite{Braunstein1994,Nagaoka1989}, it is also not directly linked to the best possible performance over the set of quantum estimation strategies.
Notice that both the Fisher information and the quantum Fisher information are still well-defined even at the point $\bar \theta$ where the rank changes.
They could, however, develop a discontinuity there.

\subsection{Discontinuity of classical and quantum Fisher information}  
Suppose now that the statistical model $p_\theta$ describes the p.m.f. of a discrete random variable $X$ and that, as $\theta\to \bar\theta$, the probability $p_\theta(\bar y)$ of one of its outcomes $\bar y\in \mathcal X$ goes to zero. Since the Fisher information is computed only on the support of the model, it follows that
\be
\Delta \mathcal{F} = \lim_{\theta\to \bar \theta} \mathcal F_\theta - \mathcal F_{\bar \theta} = \lim_{\theta\to \bar \theta} \frac{[\partial_\theta p_\theta(\bar y)]^2}{p_\theta(\bar y)}.
\ee
If the limit on the RHS is non-zero, then the Fisher information is discontinuous. 

\begin{prop}
The Fisher information $\mathcal F_\theta$ at $\theta=\bar \theta$ is continuous if both the speed $v=\lim_{\theta\to \bar \theta}\partial_\theta p_\theta(y)$ and the acceleration $a=\lim_{\theta\to \bar \theta}\partial_\theta^2 p_\theta(y)$ with which $p_\theta(y)\to 0$ are zero. Otherwise, if $v=0$ but $a\neq 0$, the discontinuity is equal to $\Delta\mathcal{F} = 2 a$ and if $v \neq 0$ there is a discontinuity of the second kind.  
\begin{proof}
Follows from L'H\^{o}pital's rule.
\end{proof}
\end{prop}

We now move to the quantum case and suppose that the rank of the quantum statistical model $\rho_\theta$ diminishes by one at $\theta = \bar \theta$ because one of its eigenvalues $\lambda_{m,\theta}$ vanishes as $\theta \to \bar\theta$. Is the quantum Fisher information discontinuous as $\theta \to \bar \theta$?  By looking at the formula in Eq.~\eqref{eq:formulaQFI},  one sees that the discontinuity can be evaluated as (in the following we are omitting the dependence on $\theta$ of eigenvalues and eigenvectors) 
\begin{align}
\Delta\mathcal{Q} &= \lim_{\theta\to \bar \theta} \mathcal Q_\theta - \mathcal Q_{\bar \theta}  \\
&=  \lim_{\theta\to \bar \theta} \left(4 \sum_{\tiny{\lambda_{k}=0}} \frac{ | \langle \lambda_{k} | \partial_\theta \rho_\theta | \lambda_{m}\rangle |^2 }{\lambda_{m}} + 2\, \frac{ | \langle \lambda_{m} | \partial_\theta \rho_\theta | \lambda_{m}\rangle |^2 }{2 \lambda_{m}} \,.
\right)
\end{align}
By analyzing the first term, where the sum runs over the kernel of the statistical model $\rho_\theta$, we observe
\begin{align}
 \frac{ | \langle \lambda_{k} | \partial_\theta \rho_\theta | \lambda_{m}\rangle |^2 }{\lambda_{m}} &= \frac{1}{\lambda_m}  | \langle \lambda_k | \partial_\theta \lambda_m \rangle |^2 \, | \langle \lambda_m |  \lambda_m \rangle |^2 \, \lambda_m^2  \nonumber \\
 &= \lambda_m | \langle \lambda_k | \partial_\theta \lambda_m \rangle |^2 \underset {\theta \to \bar\theta}{\to} 0 \,,
\end{align}
as, by hypothesis, $\displaystyle \lim_{\theta \to \bar\theta}  \lambda_m = 0$. The second term, by exploiting 
the orthogonality of the eigenstates of $\rho_\theta$, reads
\begin{align}
\frac{ | \langle \lambda_{m} | \partial_\theta \rho_\theta | \lambda_{m}\rangle |^2 }{ \lambda_{m}} &= 
\frac{ | \langle \lambda_{m} | \partial_\theta (\lambda_m |\lambda_m\rangle\langle \lambda_m|)  | \lambda_{m}\rangle |^2 }{ \lambda_{m}} \nonumber \\
&= \frac{(\partial_\theta \lambda_m)^2}{\lambda_m} + 2 \lambda_m |\langle \partial_\theta \lambda_m | \lambda_m\rangle|^2  \underset {\theta \to \bar\theta}{\to} \frac{(\partial_\theta \lambda_m)^2}{\lambda_m} \,.
\end{align}
We are thus left with the following proposition:
\begin{prop}
\label{prop:4}
The quantum Fisher information $\mathcal Q_\theta$ at $\theta=\bar \theta$ is continuous if both the speed $v=\lim_{\theta\to \bar \theta}\partial_\theta 
\lambda_m$ and the acceleration $a=\lim_{\theta\to \bar \theta}\partial_\theta^2 \lambda_m$ with which the eigenvalue $\lambda_m$ is vanishing are zero. Otherwise, 
if $v=0$ but $a\neq 0$, the discontinuity is equal to $\Delta \mathcal{Q}=2 a$
 and if $v \neq 0$ there is a discontinuity of the second kind.  
\end{prop}
The discontinuity of the QFI in variable-rank models have been addressed in \cite{Safranek2017a}, where in particular it was shown that the continuous version of the standard QFI is proportional to the Bures metric, i.e.
\be
\lim_{\theta \to \bar\theta} \mathcal{Q}_\theta  =4 g_\theta  \,.
\ee
This means that the Fisher-Bures metric can be exploited to evaluate the QFI
only for regular models whereas for non regular ones this link is broken.
In addition, as we pointed out above, the hypotheses at the basis of the 
derivation of the classical and quantum CRBs do not hold 
for this kind of models, and consequently these bounds can be violated and 
are of no use in quantum metrology.
To better describe this issue, in the next section we provide two simple quantum estimation problems falling into this class of models.
In both cases, we will show that the QFI is in fact discontinuous and, in turn, it is easy to construct an estimator with zero variance in the case $\theta=\bar\theta$.
The existence of zero-variance estimators for non-regular models with parameter-dependent support is well-known in classical estimation~\cite{Akahira1995}.

Finally, we mention that the continuity of the QFI as a functional of the operators $\rho$ and $\partial_\theta \rho$ has been studied~\cite{Rezakhani2015}.
Our point of view, similarly to~\cite{Safranek2017a}, is to investigate discontinuities of the QFI as a function of the parameter itself, by assuming that $\rho$ and $\partial_\theta \rho$ are both continuous functions of $\theta$, also at the value $\bar\theta$.
Nonetheless, the results of~\cite{Rezakhani2015} seem to be consistent with our findings.
\section{Examples of quantum statistical models with parameter dependent rank}
Let us now illustrate two paradigmatic examples of variable-rank quantum
statistical models: a model that can be mapped to a classical statistical
model, such as the ones discussed in~\cite{Safranek2017a}, and a genuinely \emph{quantum} statistical model.
For the sake of clarity, in both the examples we consider single-qubit systems.

We will show that the maximum likelihood estimator has zero variance when $\theta=\bar\theta$; this is an example of the meaninglessness of the CRB at this critical true value of the parameter.
However, we do not know if it is possible to build such zero-variance estimators for all variable-rank quantum statistical models.
\subsection{A \emph{classical} quantum statistical model}
A quantum statistical model is said to be \emph{classical} if the family of quantum states $\rho_\theta$ can be diagonalized with a $\theta$-independent unitary, and thus the whole information on the parameter is contained in the eigenvalues~\cite{Suzuki2018}.\\
The simplest example of classical model is described by the family of two-dimensional quantum states
\begin{align}
\label{eq:pparam}
\rho_p = p |0\rangle\langle 0| + (1-p) |1\rangle\langle 1| \,, \,\,\,\,\, 0\leq p \leq 1 \,.
\end{align} 
As it is apparent, this is also a variable-rank statistical model, when the parameter to be estimated $p$ takes the limiting values $\bar{p}=\{0,1\}$.
For a generic value of the parameter between the two limiting values, $0<p<1$, the QFI reads
\begin{align}
\mathcal{Q}_p = \frac{1}{p(1-p)} \,,
\end{align}
while in the two limiting values $\bar{p}$, one gets $\mathcal{Q}_{\bar{p}} = 1$. As expected, one observes a discontinuity of the second kind, and in particular an infinite Bures metric, $\lim_{p \rightarrow \bar p}  g_p = \infty$.

The optimal measurement corresponds trivially to the projections on the states $\{|0\rangle\langle0| , |1\rangle\langle 1|\}$ and clearly does not depend on the parameter to be estimated.
For the limiting values, it is easy to check that a maximum likelihood estimator would give a variance equal to zero (in fact one always gets the same measurement outcome). 
It is well known that the estimation of parameters on the boundary of the parameter space breaks down the asymptotic normality of the maximum-likelihood estimator as well as the validity of the CRB~\cite{Chernoff1954,Moran1971,Self1987,Andrews1999,Davison2003}.

Therefore, while on the one hand it should be now clear that no CRB holds in these instances, on the other hand this result would induce to say that the Bures metric gives the correct figure of merit to asses the performances for $\theta\to \bar\theta$.
However, this is not always the case: one could check that by reparametrizing the family of states to~\cite{Safranek2017a}
\be
\label{eq:thetaparam}
\rho_\theta = \sin^2 \theta | 0 \rangle\langle 0| + \cos^2 \theta |1\rangle\langle 1| \,,
\ee
one would obtain that the Bures metric is identically equal for all values of $\theta$, $g_\theta = 1$, while the standard QFI is discontinuous and reads
\begin{align}
\mathcal{Q}_{\theta} = \left\{
\begin{array}{c}
4 \,\,\,\, \theta \neq k \pi /2 \\
0 \,\,\,\, \theta = k \pi /2 \\
\end{array}
\right. 
\end{align}
However, also in this case the optimal measurement and the maximum likelihood estimator would trivially give a zero variance estimation (at least if we restrict the values of $\theta$ to $[0, \pi/2]$, so that the model becomes identifiable). In turn, the CRB is violated.

\subsection{A \emph{genuine} quantum statistical model}
Let us now consider the quantum statistical model described by a family of two-dimensional quantum states $\rho_\theta$ that solve the Markovian master equation
\begin{align}
\frac{d\rho_\theta}{dt} = - i \frac{\theta}{2} [ \sigma_z , \rho_\theta ] + \frac{\kappa}{2} ( \sigma_x \rho_\theta \sigma_x - \rho_\theta) \,,
\label{eq:ME}
\end{align}
with initial condition $\rho_\theta (t=0) = | + \rangle \langle + |$, where $|+\rangle = (|0\rangle+|1\rangle)/\sqrt{2}$ denotes the eigenstate of the $x$-Pauli matrix $\sigma_x$ in terms of the eigenstates of the z-Pauli matrix $\sigma_z$. From a physical point of view this master equation describes the evolution of a spin-$1/2$ system, subjected to a phase-rotation due to a magnetic field proportional to $\theta$ along the $z$-direction and subjected to {\em transverse} noise along the $x$-direction with rate $\kappa$.

The master equation can be solved analytically and the corresponding QFI $\mathcal{Q}_\theta$ and Bures metric $g_\theta$ can be readily evaluated.
For the parameter value $\bar\theta = 0$, one observes how the master equation has no effect on the initial state $|+\rangle$ (the Pauli matrix $\sigma_x$ clearly commutes with its eigenstate $|+\rangle\langle +|$).
As a consequence, for $\bar\theta=0$, the quantum state remains pure (and identical to the initial state) during the whole evolution, showing that the rank of the corresponding quantum statistical model changes by considering a non-zero frequency $\theta \neq 0$. Remarkably, unlike the previous example, both eigenstates and eigenvalues of $\rho_\theta$ depend on the parameter $\theta$; in this sense, the quantum statistical model cannot be readily mapped onto a classical one.

The QFI and Bures metric at the discontinuity point $\theta=\bar\theta$ can be analytically evaluated as 
\begin{align}
\mathcal{Q}_{\bar \theta = 0} &= 4 \frac{ e^{-\kappa t} \sinh^2 \frac{\kappa t}{2}}{\kappa^2}
\, \\
g_{\bar \theta =0 } &= 2 \frac{e^{-\kappa t}  + \kappa t - 1}{\kappa^2} \,.
\end{align}
One can check that the eigenvalue of $\rho_\theta$ that goes to zero for $\theta \rightarrow 0$ is equal to
\begin{align}
\lambda = \frac{1}{2} \left( 1 + \frac{e^{-\frac{\kappa t}{2}} \sqrt{\kappa^2 \cosh (\xi t) + \kappa \xi  \sinh (\xi t) - 4 \theta^2}}{\xi} \right) \,,
\end{align}
where $\xi = \sqrt{\kappa^2 - 4\theta^2}$.
The corresponding \emph{speed} and \emph{acceleration} read
\begin{align}
v &= \lim_{\theta \rightarrow 0} \partial_\theta \lambda =0 \,, \\
a &= \lim_{\theta \rightarrow 0} \partial_\theta^2 \lambda = \frac{2 \kappa t + 4 e^{-\kappa t} - e^{-2 \kappa t} + 3}{\kappa^2} \,.
\end{align}
It is then easy to check that, as predicted by Proposition~\ref{prop:4}, one gets $\Delta\mathcal{Q} = 2 a$. 

If one performs the optimal measurement for $\theta=0$, that trivially corresponds to the projection on eigenstates of the Pauli operator $\sigma_x$, one can build a maximum likelihood estimator yielding a zero variance.
However it is important to remark that, contrarily to the previous example, in this model, the optimal measurement generally depends on the true value of the parameter $\theta$, and it has to be implemented via an adaptive strategy, as previously mentioned.
The adaptive scheme will be equivalent to the optimal measurement only asymptotically, while for arbitrarily large but finite number of repetitions $M$, one will implement a strategy with a small but finite difference from the optimal one.
In such a scenario, we heuristically expect that the variance of the \emph{asymptotic} estimator will be bounded by (four times) the \emph{continuous} Bures metric $g_{\bar\theta}$, obtained via the limit of $\theta$ going to $\bar{\theta}$, and that one should consider this figure of merit to quantify the overall performance of the estimation process.
While we lack a rigorous proof of this intuition, we believe that studying the performances of realistic adaptive estimation schemes for these non-regular models is an open topic for future research.

It is worth to remark that this quantum statistical model can be generalized to $N$ qubits.
In~\ref{app:GHZ}, for an initial GHZ state, we evaluate the limit of the QFI for $\theta \to 0$, i.e. the Bures metric in $\theta=0$, as well as the discontinuous QFI obtained for $\theta=0$ and we underline a markedly different behavior of the two quantities as functions the probing time $t$.
Incidentally, the limit of the QFI for $\theta \to 0$ is exactly the quantity that we dubbed \emph{ultimate} QFI for continuously monitored quantum systems, i.e. obtained by optimizing over all the possible measurements on the system and the environment causing the Markovian non-unitary dynamics~\cite{Gammelmark2014,Albarelli2018a,Albarelli2018}.
In such a framework, this quantity represents a valid statistical bound for all values of $\theta$.
\section{Conclusion}
The quantum Cram\'er-Rao theorem is regarded as the foundation of quantum estimation theory, promoting the QFI and the Bures metric as the two fundamental figures of merit that one should consider in order to obtain the ultimate precision achievable in the estimation of parameters in quantum systems.
In this manuscript we have addressed variable-rank quantum statistical models and the corresponding discontinuity of the QFI.
While this topic has been addressed before in the literature~\cite{Safranek2017a}, the validity of the quantum Cram\'er-Rao theorem was not properly discussed. 
Here, we have shown in detail that the proof of the theorem in fact breaks down in these pathological cases both in the classical and quantum case; as a consequence, the bound can be no longer considered valid.
We have also addressed two paradigmatic examples, and considered the corresponding behaviour of the QFI, of the Bures metric and of the variance of the maximum likelihood estimator.
While in regular cases these quantities typically coincide, we have shown how they may differ in the presence of a discontinuity.
In particular, for these qubit examples the maximum likelihood estimator is actually deterministic at the critical true value of the paramater, but we do not know wether this is true for variable-rank models in general.
Overall our results, apart from contributing to clarifying previous results on the discontinuity of the QFI, pave the way to further studies on the relationships between optimal estimators and Bures metric in non-regular quantum statistical models.
%
\ack
We acknowledge stimulating discussions with D.~Branford, M.~A.~C.~Rossi, D.~\v{S}afr{\'{a}nek, A.~Smirne, D.~Tamascelli and T.~Tufarelli.
This work has been partially supported by JSPS through FY2017 program (grant S17118).
FA acknowledges support from the UK National Quantum Technologies Programme (EP/M013243/1).
MGG acknowledges support from a Rita Levi-Montalcini fellowship of MIUR.
MGAP is member of GNFM-INdAM. 
\appendix
\section{Discontinuity for frequency estimation with N-qubit GHZ states in transverse independent noise}
\label{app:GHZ}

The calculation presented in this Appendix is hinted, but not included, in Ref.~\cite{Albarelli2018a}; we also mention that frequency estimation with transverse noise and a vanishing parameter was also considered in Appendix D of~\cite{Brask2015}, but without highlighting the appearance of a discontinuity.

\subsection{Evolution of a GHZ state in transverse noise}
Greenberg-Horne-Zeilinger (GHZ) states are the prototypical example of states showing a quantum advantage in metrology.
Furthermore, they have a particularly simple evolution under independent noises acting on the qubits~\cite{Aolita2008,Aolita2009}.
In particular, we consider a $N$-qubit GHZ state 
\begin{equation}
|\psi_\mathsf{GHZ}\rangle = ( |0\rangle^{\otimes N} + |1\rangle^{\otimes N})/\sqrt{2} \;,
\end{equation}
evolving according to a $N$-qubit version of the master equation~\eqref{eq:ME}:
\begin{equation}
\label{eq:MarkovFreq}
\frac{d\rho}{dt} =  -i \frac{\theta}{2} \sum_j^N \left[ \sigma_z^{(j)},\rho \right] + \frac\kappa{2} \left( \sum_{j=1}^N \sigma_x^{(j)} \rho \sigma_x^{(j)} - N \rho \right) \;,
\end{equation}
where the the superscript $(j)$ labels operators acting on the $j$-th qubit (i.e. tensored with the identity on all the other qubits).
The evolved state $\rho$ becomes a mixture of states of the form $|s \rangle \pm | \bar{s} \rangle$, where $s$ is a binary string and $\bar{s}$ is its bitwise negation, e.g $\ket{s} = \ket{00101}$ and $\ket{\bar{s}} = \ket{11010}$.
In the computational basis the density matrix maintains a cross-diagonal form.

It is clever to parametrise the matrix elements with an index $m \in [0,N]$, which counts how many $1$s appear in the binary string $s$, i.e. the sum of the binary representation of $s$.
Since we have $N$ qubits there are $2^N$ different possible strings, and there are $\binom{N}{m}$ different binary strings that sum to the value $m$, so that $\sum_{m=0}^{N} \binom{N}{m} = 2^N$.
It turns out that the matrix elements of an evolved GHZ state only depend the value $m$.
With such a parametrization we have the following matrix elements~\cite{Chaves2013}
\begin{equation}
\label{eq:GHZtransnoise_matel}
\begin{split}
\rho_{m,m} &= \frac{1}{2} \left[ d^m a^{N-m} + d^{N-m} a^m \right] \\ 
\rho_{m,N-m} &= \frac{1}{2} \left[ f^m \left( b - i c \right)^{N-m} + f^{N-m} \left( b + i c \right)^m \right],
\end{split}
\end{equation}
where we can further notice the symmetry of the diagonal terms under the exchange $m \to N - m$.
The coefficients appearing in the expression are given by
\begin{align}
a = \frac{1}{2}  \left( 1 + e^{-\kappa  t} \right) \qquad
d = \frac{1}{2}  \left( 1 - e^{-\kappa  t} \right) \qquad b = e^{-\frac{\kappa  t}{2}} \cosh \left(\frac{t}{2} \sqrt{\kappa^2-4 \theta^2}\right) \\ 
f 	= \kappa \frac{e^{-\frac{\kappa t}{2}} \sinh \left(\frac{t}{2} \sqrt{\kappa^2-4 \theta^2}\right)}{ \sqrt{\kappa^2-4 \theta^2} } 
\qquad c  = 2 \theta \frac{ e^{-\frac{\kappa t}{2}} \sinh \left(\frac{t}{2} \sqrt{\kappa^2-4 \theta^2}\right) }{ \sqrt{\kappa^2-4 \theta^2} }. \nonumber
\end{align}
All these coefficients are real as long as $\theta < \frac{\kappa}{2}$, which is the case we are interested in, since we want to take the limit $\theta \to 0$.

\subsubsection{Piecewise QFI for a qubit}

The QFI for qubit states can be very conveniently written via the Bloch representation of qubit states~\cite{Zhong2013}:
\begin{equation}
\rho = \frac{1}{2} \left( \id + \vec{v} \cdot \vec{\sigma} \right);
\end{equation}
the QFI is easily expressed in terms of the Bloch vector $\vec{v}$
\begin{equation}
\label{eq:QFIqubit}
\mathcal{Q}_\theta \left[ \rho \right] = 
\begin{cases}
\left| \partial_\theta \vec{v} \right|^2 + \frac{|\partial_\theta \vec{v} \cdot \vec{v} |^2 }{1 - | \vec{v}|^2 } \qquad &|\vec{v}| < 1  \\ 
\left| \partial_\theta \vec{v} \right|^2  &|\vec{v}| = 1.
\end{cases}
\end{equation}
From this piecewise definition it is easy to see the possibility of a discontinuity.
In particular, for a qubit the only possible change of rank is that for $\bar{\theta}$ the state becomes pure, i.e. $|\vec{v}| = 1$ and so the $\lim_{\theta \to \bar{\theta}}$ gives rise to a $\frac{0}{0}$ indeterminate form; we can then use L'H\^{o}pital's rule and get
\begin{equation}
\label{eq:QFIqubit_limit}
\lim_{\theta \to \bar{\theta}} \mathcal{Q}_\theta \left[ \rho \right] = - \vec{v} \cdot \partial^2_\theta \vec{v} |_{\theta=\bar{\theta}}.
\end{equation}

\subsection{Continuous and discontinuous QFI}
Given the cross structure of the evolved density matrix and the symmetry of the elements~\eqref{eq:GHZtransnoise_matel}, we can reshuffle the $2^{N} \times 2^N$ density matrix and write it as the direct sum of $2 \times 2$ matrices defined as follows
\begin{equation}
\varsigma_m = \begin{pmatrix}
\rho_{m,m} & \rho_{m,N-m}\\
\rho_{m,N-m}^* & \rho_{m,m}\\
\end{pmatrix},
\end{equation}
where now we need only half the values of the index $m=0,\dots,\lfloor N/2 \rfloor$.
Each of these $\varsigma_m$ is repeated $\binom{N}{m}$ times, except the last matrix for $m=\lfloor N/2 \rfloor$ that appears $\frac{1}{2} \binom{N}{m}$ times if $N$ is even and $\binom{N}{m}$ times if $N$ is odd.
This reshuffling is obtained by applying orthogonal permutation matrices that do not change the QFI.
The diagonal elements do not depend on $\theta$ and the derivative of $\varsigma_m$ reads
\begin{equation}
\partial_\theta \varsigma_m = \begin{pmatrix}
0 & \partial_\theta \rho_{m,N-m}\\
\partial_\theta \rho_{m,N-m}^* & 0\\
\end{pmatrix}.
\end{equation}
We can renormalized the matrices $\varsigma_m$ to get proper qubit states, i.e.
\begin{equation}
\tilde{\varsigma}_m = \frac{1}{2} \begin{pmatrix}
1 & \frac{\rho_{m,N-m}}{\rho_{m,m} } \\
\frac{\rho_{m,N-m}^*}{\rho_{m,m}} & 1\\
\end{pmatrix}
\, \qquad \,
\partial_\theta \tilde{\varsigma}_m = \frac{1}{2} \begin{pmatrix}
0 & \frac{\partial_\theta \rho_{m,N-m}}{\rho_{m,m} } \\
\frac{\partial_\theta \rho_{m,N-m}^*}{\rho_{m,m}} & 0\\
\end{pmatrix}.
\end{equation}
With this normalization the Bloch vector of $\tilde{\varsigma}_m$ is then $\left[ \mathrm{Re} \left( \rho_{m,N-m} \right), \mathrm{Im} \left(\rho_{m,N-m} \right), 0 \right] / \rho_{m,m}$.

It is not hard to see that the QFI of the global state is the average of the QFIs of these qubit states
\begin{equation}
\label{eq:QFI_GHZ_qubit}
\mathcal{Q} \left[ \rho \right] = \sum_{m=0}^{N} \binom{N}{m}  \rho_{m,m} \mathcal{Q} \left[ \tilde{\varsigma}_m \right] ,
\end{equation}
where the factor $2$ in the normalization vanishes because we have extended the sum to $N$ and divided by $2$, this is possible since $\mathcal{Q} \left[ \varsigma_m \right]=\mathcal{Q} \left[ \varsigma_{N-m} \right]$ (the two states differ only for a conjugation of the off-diagonal elements).
For $\theta = 0 $ we have that $\rho_{m,N-m} = \rho_{m,m}$ and the states $\tilde{\varsigma}_m$ all become pure, with Bloch vector $[1,0,0]$, so that the global $N$ qubit state goes from full rank (i.e. $2^N$) to rank $2^{N-1}$.
To calculate the limit of the QFI for $\theta \to 0$ we need to use Eq.~\eqref{eq:QFIqubit_limit}.
In this case it is possible to compute the sum~\eqref{eq:QFI_GHZ_qubit} explicitly
\begin{equation}
\mathcal{Q}_{\theta \to 0} =   -\sum_{m=0}^{N} \binom{N}{m} \partial^2_\theta \rho_{m,N-m} \Bigg|_{\theta=0} = \frac{ N^2 \left(1- e^{-\kappa  t}\right)^2 + N \left[ 2 \kappa  t+ 1 - \left(2 - e^{-\kappa t} \right)^2 \right] }{\kappa ^2} ,
\end{equation}
this equation corresponds to the ultimate QFI obtained in~\cite{Albarelli2018a}, i.e. the best possible precision achievable by continuously measuring the environment degrees of freedom causing the non-unitary part of the Markovian evolution~\eqref{eq:MarkovFreq}.

On the other hand, the discontinuous QFI for $\theta = 0$ is obtained by applying the second line of Eq.~\eqref{eq:QFIqubit}, which results in
\begin{equation}
\mathcal{Q}_{\theta = 0} = \sum_{m=0}^{N} \binom{N}{m} \frac{\left\vert \partial_\theta \rho_{m,N-m} \right\vert^2}{\rho_{m,m}} \Bigg|_{\theta=0} .
\end{equation}
In Fig.~\ref{fig:timecontVSdiscont} we plot the two quantities $\mathcal{Q}_{\theta \to 0}$ and $\mathcal{Q}_{\theta = 0}$ for some values of $N$ as a function of the evolution time $t$.
From the plots one can see that the behaviour for long evolution times is indeed dramatically different.

\begin{figure}[h!]
	\centering
\begin{subfigure}[b]{0.4\textwidth}
	\centering
	\includegraphics[width=\textwidth]{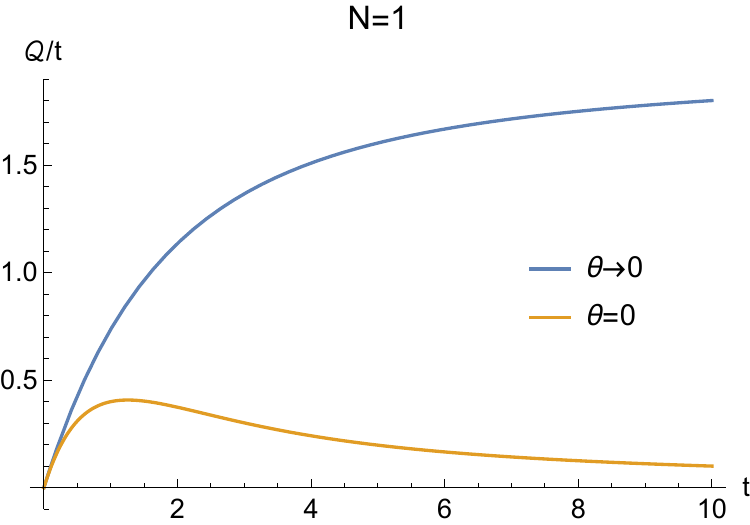}
	\caption{}
	\label{}
\end{subfigure}
\begin{subfigure}[b]{0.4\textwidth}
	\centering
	\includegraphics[width=\textwidth]{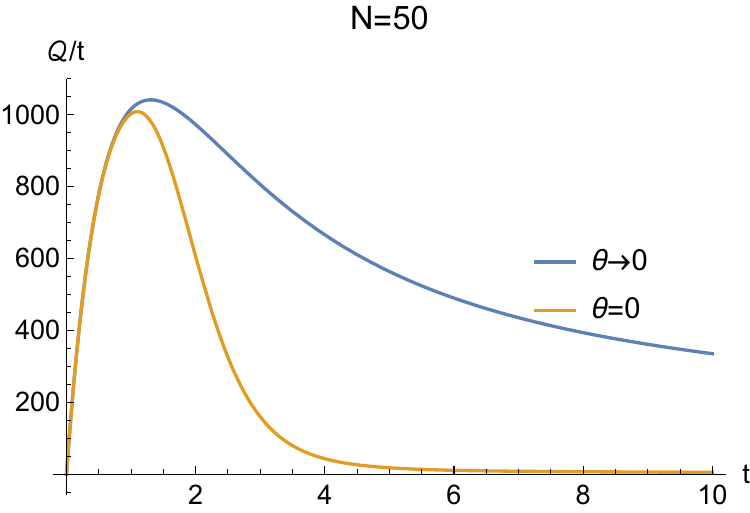}
	\caption{}
	\label{}
\end{subfigure}
\vskip\baselineskip
\begin{subfigure}[b]{0.4\textwidth}
	\centering
	\includegraphics[width=\textwidth]{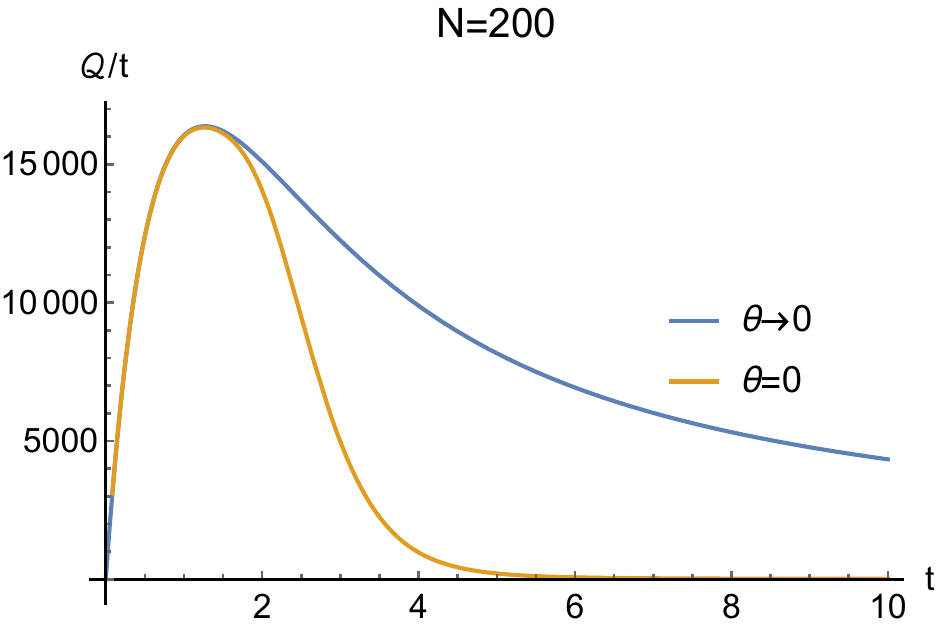}
	\caption{}
	\label{}
\end{subfigure}
\caption{Plots of the QFI per unit time as a function of time, for $\kappa=1$, for different values of $N$.}%
\label{fig:timecontVSdiscont}%
\end{figure}
\newpage

\section*{References}
\bibliographystyle{iopart-num}
\bibliography{vrbiblio}
\end{document}